\documentstyle[prl,aps,psfig,floats,twocolumn]{revtex}

\def\FC{fundamental cluster}
\def\FCs{fundamental clusters}
\def\MF{mean-field}
\def\NMF{near-mean-field}
\input epsf.sty

\begin{document}

\twocolumn[\hsize\textwidth\columnwidth\hsize\csname
@twocolumnfalse\endcsname

\title{Scaling in a cellular automaton model of earthquake faults}

\author{M.\ Anghel,${}^{\dag\pm}$ W.\ Klein,${}^{\dag\pm}$, 
  J.\ B.\ Rundle,${}^{\ddag\star}$ and J.\ S.\ S\'a
Martins${}^{\star}$}

\address{${}^{\dag}$Center for Nonlinear Studies, LANL, Los Alamos,
NM 87545}

\address{${}^{\ddag}$Physics Department, University of Colorado at
Boulder, Boulder, CO 80309}

\address{${}^{\star}$Colorado Center for Chaos and Complexity, CIRES, 
University of Colorado at Boulder, Boulder, CO 80309}

\address{${}^{\pm}$Permanent address: Physics Department and Center 
for Computational Science, Boston University,  Boston, MA 02215}

\maketitle

\begin{abstract}
  We present theoretical arguments and simulation data indicating that 
  the scaling of earthquake events in
  models of faults with long-range stress transfer is composed of at
  least three distinct regions. These 
  regions correspond to three classes of earthquakes with different 
  underlying physical mechanisms. In addition to the 
  events that exhibit scaling, there are larger ``breakout'' events 
  that are not on the scaling plot. We 
  discuss the interpretation of these events as fluctuations in the
  vicinity of a spinodal critical point. 
\end{abstract}


]

Earthquake faults and fault systems are known to
exhibit scaling\cite{gr,scholz} where
the number $N_{M}$ of earthquakes with seismic moment $M$  scales as
$N_{M}\sim 1/M^{B}$ with $B$ between 1.5 and 2.0\cite{scholz}.
The observed scaling is over several decades, but for the larger
events  there 
is an indication that scaling does not apply, a fact
often attributed to poor statistics. However, because models also 
produce this deviation from scaling, even when there are 
many large events\cite{langer,ferg}, the origin of this deviation
lies elsewhere.
Other questions of interest include: What is the physical mechanism 
that produces the scaling? Do all the events on the scaling
plots have the same physical origin? We report
the results of our theoretical and numerical investigations of a cellular 
automaton (CA) model of an earthquake fault 
indicating that the scaling region is dominated by a spinodal-like 
(pseudospinodal) singularity that determines the
distribution of events. The scaling can be decomposed into three distinct
regions driven by different physical mechanisms. In addition to the 
scaling region, we find that the largest  ``earthquakes'' 
are not on the scaling plot and have yet another physical origin. 

The system of interest is a CA version of the slider block
model\cite{bk}  and consists of a discrete two-dimensional ($d=2$) 
array of blocks connected by
linear springs with a spring constant (stress Green's function)
$T(r_{ij})$ and to a loader plate by linear springs with
constant $K_{L}$; $r_{ij}$ is the
distance between blocks. Each block $i$ initially receives a
random  position $U_{i}$ from a uniform distribution, and the loader
plate  contribution to the stress is  set to
0. The stress $\sigma_{i}$ on each block is given by 
$\sigma_{i}(t) = \sum_{j}T(r_{ij})[U_{j}(t) - U_{i}(t)] + 
K_{L} [V \sum_{n}\Theta(n-t) - U_{i}(t)]$ and
compared  to a threshold value $\sigma^{F}_{i}$. If
$\sigma_{i}<\sigma^{F}_{i}$, the block is not moved. If 
$\sigma_{i}\geq \sigma^{F}_{i}$, the block slips (fails) and is moved
according to
$U_{i}(t+1) = U_{i}(t) + 
[\sigma_{i}(t) - \sigma_{i}^{R}(t)]/K,$
where $K=K_{L}+K_{C}$, and $K_{C}=\sum_{j,i\neq j}T(r_{ij})$. 
The residual stress, 
$\sigma_{i}^{R}(t)= \sigma^{R}+ a (\eta_{i}(t)-0.5)$,
specifies the stress on a block immediately after 
failure. The random noise $\eta_{i}$ is taken from a 
uniform distribution between $0$ and $1$, $a$ sets the noise
amplitude, and $\sigma^{R}$ is the average residual stress.
After all the  blocks have been tested and moved, the 
stress on each block is measured again
and the process is repeated. We
choose $T_{ij}=K_{C}/q$ for all
$j$ inside a square  interaction range with area $(2R+1)^{2}$ 
centered on site $i$, where $q =(2R+1)^{2}-1$ is the number
of neighbors; $T_{ij}=0$
for all the sites outside the interaction range. After block $i$
slips,
$K_{C}/K$ of the local stress drop, $\sigma_{i} -
\sigma_{i}^{R}$, is distributed equally to its
neighbors, and $K_{L}/K$ is dissipated.
When no block
has a stress greater than $\sigma_{i}^{F}$, the earthquake ceases
and the seismic moment released during the event is
$M = \sum_{i} \Delta U_{i} $, where $\Delta
U_{i}$ is the slip of block $i$ during the earthquake. 
The loader plate is then moved a distance $V\Delta
T$, the stresses are updated, and we search for the unstable 
blocks that will initiate the 
next event. The quantity $\Delta T,$ which we set equal to 1, 
sets the ``tectonic'' time scale. In the limit $V=0$ the stress
is globally incremented to bring the ``weakest'' block to failure and
there is a single initiator per plate update. 

\begin{figure}[h!]
  \hbox to\hsize{\epsfxsize=1.0\hsize
    \hfill\epsfbox{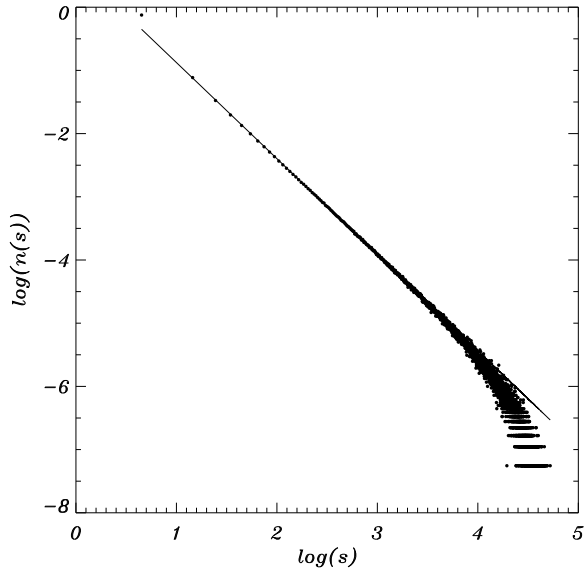}\hfill}
  \caption{Log-log plot of the probability $n(s)$, the number
    of events  with $s$ failing blocks divided by their total
number, $18 \times  10^{6}$. The system consists of $256 \times 256$
blocks with periodic boundary conditions.
    Note the deviation from the straight line of slope  $1.5\pm
0.05$ for large $s$.}
  \label{all_events_scaling}
\end{figure}

Because the $T_{ij}$ appropriate for earthquake
faults is long-range\cite{ferg}, 
we will consider $R >> 1$.
In our simulations $R=30$, $\sigma^{F}_{i}=\sigma^{F}=1$ is  a
spatial constant, $K_{L}=1$, $K_{C} = 100$, $V=0$, and the distribution of
residual stresses is defined by $\sigma^{R}=0.25$ and $a = 0.5$. 
In Fig.~\ref{all_events_scaling} we plot the  log (base 10) of the
probability $n(s)$ of events of size $s$ (number of failing blocks)
versus $\log(s)$  generated by the model. For the chosen
parameters there are no multiple failures of the same block during
an earthquake and $M \sim s$. For the total of
$18 \times 10^{6}$ events, there is still a significant 
spread of the data in the large events region. The origin of this
spread is not poor statistics.

We now review the theoretical arguments that describe the scaling
of events. In the limit that $R$ diverges such that 
$\!\int\! r^{2}T(r)d{\vec r}\rightarrow \infty$ but $\!\int\!
T(r)d{\vec r}$ is finite, we have derived a Langevin equation for
the stress\cite{ferg,kl}. This derivation and numerical
simulations\cite{ferg,run} confirm that the CA model is described
by equilibrium statistical mechanics in the limit of
$R \to \infty$ and that this description is a very good
approximation for systems with long, but not infinite, $R$. Because
this Langevin equation is a general description of systems with a
simple scalar order  parameter\cite{ma}, the scaling of the
fluctuations in the vicinity of  spinodals of  mean-field
Ising models (and simple fluids) and the present CA model  is the
same.

Our main assumption is  that the  structure and  dynamics of
earthquake events is identical to the structure and dynamics of
fluctuations near critical points and  spinodals. In particular,
scaling  is determined by the presence of  a spinodal
singularity\cite{ferg,kl}. For Ising systems, by mapping 
the thermal critical point onto a properly chosen
percolation  model\cite{ck,klein}, the properties of
the fluctuations at the  thermal critical point  can be obtained
from the properties of the clusters at the percolation threshold: 
percolation clusters are the physical realization of the
fluctuations\cite{ck,klein}. At  the critical point the clusters
associated with the divergent connectedness length are the
fluctuations associated  with the divergent susceptibility in the
thermal model. We can use a similar  mapping to generate a
percolation model for the spinodal. Therefore,  we can describe
the scaling of events in the  CA model  in the language of cluster
scaling for Ising models.

We first discuss how the cluster structure relates to thermal
critical phenomena in non-mean-field systems ($R=1$) where
hyperscaling is  valid.  In this case, the mean number of
clusters in a  region of volume $\xi^{d}$,
where $\xi$ is the correlation
length, is one. In such systems the critical phenomena fluctuation in 
this volume is isomorphic to the cluster\cite{ck}. This picture is
altered in \MF\
systems. For \MF\ Ising systems ($R \rightarrow \infty$)  there is a
line of spinodal critical points in addition to the  usual critical
point. These \MF\ thermal singularities can also be mapped onto
percolation transitions\cite{klein}, but the relation between
percolation clusters and critical fluctuations is qualitatively
different.  The mean number of clusters in a
volume $\xi^{d}$ is $N_{c}=R^{d}\epsilon^{2-d/2}$
near the critical point and $N_{s}=R^{d}\Delta h^{3/2-d/4}$ near
the spinodal\cite{ketal,mk,rk1,rk2}. Here $\epsilon=(T-T_{c})/T_{c},$
where $T_{c}$ is the critical temperature, and 
$\Delta h=h-h_{s},$ where $h_{s}$ is the value of the magnetic field at 
the spinodal for a fixed temperature $T<T_{c}.$ The factor $R^{d}$
appears because all lengths are in units of the interaction range.
The Ginsburg criterion for \MF\ critical points is
$\epsilon^{-\gamma}/(R^{d}\epsilon^{2\beta-d\nu})<<1$\cite{ma}.
That is, the system is well approximated by \MF\ theory if 
the fluctuations are small compared to the order parameter.
Using the \MF\ exponents\cite{ma}
$\gamma=1$, $\beta=1/2$ and $\nu =1/2$, 
the Ginsburg criterion is equivalent to $N_{c}>>1$. We will refer
to systems with $N_{c}>>1$ but finite as near-mean-field. A
similar argument is  used near the spinodal to show that $N_{s}>>1$
for \NMF. 

Because $N_{c}>>1$, the meaning of order parameter 
scaling is changed. For systems with hyperscaling\cite{ma}, the
density of the  single cluster with diameter $\xi$ scales as 
$\epsilon^{\beta}$, as does the order parameter. In \MF\ and
\NMF\ systems, $\epsilon^{\beta}$ cannot be the density of a
single cluster, because that would lead to a magnetization per spin
greater than one. Instead
$\epsilon^{\beta}$ is the density of all the spins in all the
clusters in a volume $\xi^{d}$\cite{ketal,mk,rk1}.  
Because all of the clusters are identical, the density of each of
these clusters is $\rho_{c}^{\rm fc}\sim
\epsilon^{1/2}/ (R^{d}\epsilon^{2-d/2})$ at \MF\ critical points
and 
$\rho_{s}^{\rm fc}\sim \Delta h^{1/2}/ (R^{d}\Delta h^{3/2-d/4})$
at spinodals. These densities are good approximations in
\NMF\ systems. We will refer to these clusters as {\it fundamental
clusters}. These clusters are not the critical phenomena
fluctuations, but are related to them\cite{ketal,mk,fja}.

Spinodals  mark the boundary between the metastable and unstable
states. In \NMF\ systems the spinodal is not a sharp singularity but
becomes a smeared out region\cite{bin} associated with
singularities in complex temperature and magnetic field
space\cite{hg}. As the spinodal is approached so is the limit of
metastability\cite{mk}. Hence, we would expect that nucleation
events, which form another class of clusters, also play a role in
the CA model.  From the Langevin
approach\cite{ferg,kl} we find that the nucleation clusters are
local regions of growth of the stable high stress phase in the
metastable low stress phase. An  earthquake represents the stress
release due to the decay of the high stress  phase into the
metastable low stress phase. Because the nucleation phenomena of 
interest occurs near the spinodal, the classical picture
is not  valid\cite{lang1,lang2}. 
Instead, a calculation of the nucleation rate  must include the
effect of the spinodal which involves a vanishing of the surface
tension\cite{uk}. With these considerations the nucleation rate, 
which is proportional to the number $n$ of clusters per unit volume, 
is given by\cite{uk}
\begin{equation}
  \label{nr}
  n \propto {\Delta h^{1/2}\exp(-AR^{d}
    \Delta h^{3/2-d/4})\over R^{d}\xi^{d}},
\end{equation}
where $A$ is a constant independent of $R$ and $\Delta h$.

The nucleation rate in Eq.~(\ref{nr}) contains an
exponential term whose argument is the nucleation barrier. The
static prefactor, which is independent of the
dynamics of the model, is $1/\xi^{d}$,
where $\xi=R \Delta h^{-1/4}$ is the correlation length near the
spinodal\cite{lang1,uk,rik}. The $\Delta h^{1/2}$ term  is the
kinetic prefactor and is dynamics specific\cite{lang2,uk}. 
For the CA model the distance from  the spinodal is measured by
the amount of stress dissipated, i.e., 
$\Delta h \sim K_{L}/K$\cite{kafrm}.
Finally, the extra
factor of $R^{d}$ in the denominator reflects the fact that the theory
employs a coarse grained time scale\cite{ferg}, but our simulations use
a time scale based on plate updates. Because the coarse graining
time is proportional to the coarse graining volume
$R^{d}$\cite{ferg2}, this extra  factor is included in the
nucleation rate. 

Our assumption is that the CA model behaves like an Ising model near the
spinodal for \MF\ and \NMF\ systems. In this
limit\cite{mk,rk1,rk2,fja}, 
\FCs\ and nucleation events, which involve coalescence of
\FCs\cite{mk}, are the only clusters. Because it is the decay of the
high stress clusters that is the ``earthquakes'' in this model, cluster
scaling and earthquake scaling are the same. To understand how
this point of view provides answers to the questions posed in the
introduction we discuss the \FCs.

In  \MF\ each block fails at the failure threshold and  
fails only once during an earthquake\cite{ferg,kl}. This behavior
is an  excellent approximation in \NMF\cite{ferg}.  The amount
$\Delta \sigma$ of stress
transmitted to a site during the failure of a cluster is 
proportional to the number of cluster sites in the
interaction volume
$R^{d}$, which is $\rho^{\rm fc}_{s}  R^{d}$ because we are
near the spinodal, times 
the fraction of  stress transmitted from a failed block,
which  is proportional to $R^{-d}$. Hence, $\Delta \sigma \sim
\rho^{\rm fc}_{s}$ during the failure of a \FC, and
the mean size of the \FC\ is $s = \rho^{\rm fc}_{s} \xi^{d}=\Delta
h^{-1}$.  The number of \FCs\  per unit volume  is 
$n_{\rm fc} = R^{d}\Delta h^{3/2-d/4}/\xi^{d} = \Delta h^{3/2}$.
Therefore, the density of \FCs\  with $s$ blocks  scales as
$n_{\rm fc}\sim 1/s^{3/2}$.

\begin{figure}[tbp]
  \hbox to\hsize{\epsfxsize=1.0\hsize
    \hfill\epsfbox{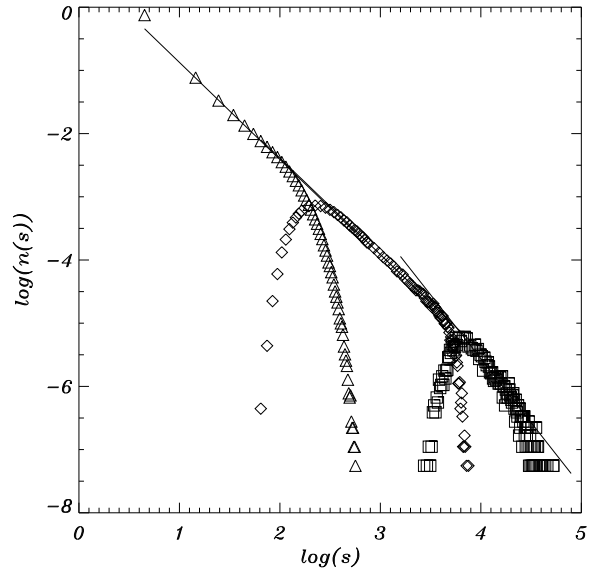}\hfill}
  \caption{Same data as Fig.~\ref{all_events_scaling}, with the
    events  separated into different classes. The 
    triangles represent \FCs\ with scaling exponent
    of $1.5 \pm 0.05$. The diamonds are failing nucleation clusters 
    with a scaling exponent  $1.5 \pm 0.05$,  showing the absence of
    critical slowing down. The open squares are arrested nucleation events
    with a scaling  exponent $2.0 \pm 0.1$ indicating
    the presence of critical slowing down. There are ``breakout''
events
    at the largest end of the plot that do not exhibit scaling.}
  \label{cluster_scaling}
\end{figure}

To identify the \FCs\  we examine the stresses on the blocks that
make up  a cluster of failed sites and determine the minimum stress,
$\sigma_{\rm min}$, of the failing sites  prior to failure, but 
after the update of the loader plate that triggers the event. We
record {\it only} those events for
which $\sigma_{\rm min}$ is within the window  
$[\sigma^{F} - \Delta \sigma, \sigma^{F}]$. In
Fig.~\ref{cluster_scaling} we plot the log of the number of \FCs\ 
versus log $s$. For the chosen parameters, $\Delta h = 0.01$ so
that
$\Delta \sigma \sim 0.01$. The slope
of 1.53 is consistent with the theoretical prediction. The mean size 
$s=\Delta h^{-1}=100$ is also consistent with our data. Note the
lack of data spread. In Figs.~\ref{all_events_scaling} and 
\ref{cluster_scaling} the \FCs\ make up only the small
$s$ end of the scaling plot, but the
\FCs\ comprise $\sim 17 \times  10^{6}$ out of 
$18 \times  10^{6}$ events. Hence, a
simulation of earthquake faults will require huge numbers of events to
probe the statistics of the interesting and important large event
region.

We now consider the  nucleation events and their clusters. The 
size of the  nucleation events depends on several factors
that determine precisely when a given high stress 
nucleation event will stop growing. We will concentrate on two
regimes. The first is events near the top of the saddle point hill 
associated with the barrier between the stable and metastable 
state\cite{lang1,uk}. The reason we neglect clusters on all scales between 
the \FCs\ and the saddle point clusters is that Ising
model studies have found no 
clusters in this intermediate region\cite{mk,fja}. Another aspect of 
nucleation events of this kind in Ising systems is that one must get
very close to the spinodal to observe critical slowing down
because the saddle point hill appears to be high but not very flat
until the system is very close to the spinodal\cite{mk,dwh}. The
absence of critical slowing down near the saddle point also has 
been seen in the CA model\cite{ferg2}. Hence, there is a class of
nucleation events that do not quite reach the top of the saddle
point hill. As a result, random fluctuations lead to the decay
of these clusters back to the metastable phase. We call these
clusters {\it failing nucleation events}. The probability of these
events is  characterized by a saddle point calculation  without the
kinetic  prefactor.  
From these considerations and Eq.~(\ref{nr}), the mean number of 
failing nucleation events per unit volume is
$n_{\rm fn}\propto \exp(-AR^{d}\Delta h^{3/2-d/4})/\xi^{d}$, the
same as  Eq.~(\ref{nr}) without the kinetic prefactor.

To obtain predictions for the scaling regime two results are needed.
The first is that $s_{\rm fn}=\Delta
h^{1/2}\xi^{d}$, where $\Delta h^{1/2}$ is the density
of the nucleating cluster\cite{uk}. 
Second, because the exponential is
a rapidly increasing function of its argument, as $\Delta h$
decreases, the probability of a cluster increases
from almost zero to a relatively large number over a very short
interval of
$\Delta h$. The value of $\Delta h$ where this crossover occurs is
the limit of metastability\cite{bin}. For this reason essentially
all of the  nucleation events take place at  a fixed value of
$AR^{d}\Delta h^{3/2-d/4}=C$. As for the \FCs\ the stress transfer
to a site in a nucleation event is equal to the density of the
event. For our parameters the density is $\Delta h^{1/2} = 0.1$.
Hence we identify these nucleating clusters by selecting {\it only} 
those events whose $\sigma_{\rm min}$ falls within the window
$[0.90, 0.99]$. The size of the event is $s_{\rm fn}=\Delta
h^{1/2}\xi^{d}\sim 1000$ and the number of these events is
$n_{\rm fn}\propto e^{C}/\xi^{d}$. Using the relation between $R$
and
$\Delta h$ implied by a fixed value of $C$, we find that
$n_{\rm fn}$ scales as $1/s^{3/2}_{\rm fn}$. In
Fig.~\ref{cluster_scaling} we show a log-log  plot of
$n_{\rm fn}$ versus
$s_{\rm fn}$. The slope and mean size are consistent with our
predictions.

Finally, we consider a second class of  nucleation
events.  These are the events that have
made it to the top and over the saddle point hill and have become
arrested during their growth phase, i.e., after growing to some
size the high stress nucleation region decays back to the low
stress  metastable state. Because the clusters have made it to the
top of the  saddle point hill, this decay of the high stress phase
is no longer induced by random fluctuations: it appears in the
Langevin approach due to a decreasing loader plate velocity on the
coarse grained scale\cite{ferg,kafrm} which pulls the system away
from the spinodal. We will call these clusters {\it arrested nucleation
events}.

Because these clusters  experience critical
slowing down, their  number per unit volume is given by
Eq.~(\ref{nr}). A key feature in the growth  of nucleation events
near the spinodal is that their initial growth is a filling
in\cite{dwh}, and hence these clusters are compact, that is
$s_{\rm an}\propto \xi^{d}=R^{d}\Delta h^{-d/4}$. The density is of
order unity so that we will identify these events with those
clusters whose  minimum stress of  the failing blocks  obeys the
condition $\sigma_{\rm min} < 0.90$.  Using the same arguments as
in for the failing  nucleation events, we find that their mean
size is about $10^{4}$ and the slope of the scaling plot is
predicted to be 2. The data presented in
Fig.~\ref{cluster_scaling} is consistent with these  predictions.

In summary, these theoretical considerations and numerical results
strongly suggest several important points. (1) Earthquake fault
models are statistically dominated by small, and in the case of
real earthquakes, uninteresting events. (2) The large and small
events have different physical mechanisms. (3) The scaling regime
is composed of events with two different power law distributions,
which accounts for the data spread at the large events end of
the scaling plot in Fig.~1. (4) Note that there is still a  spread
in the data at the large events end of Fig.~\ref{cluster_scaling}
and that these events do  not
scale with a slope of 2.  Numerical investigation indicates that
these  are ``breakout'' events that are generated by the spatial 
coalescence of arrested nucleation events\cite{AK} and are beyond
the  assumptions of our present theoretical treatment. That is, as
the arrested high stress cluster  decays, it releases stress into
the surrounding system. If, due to past history, the stress field
is unstable, this stress release can lead to runaway failure. This
type of event was considered in Ref.~\cite{run2} and is a
fourth mechanism that must be considered in the generation of
earthquakes. In contrast, the nucleation events are generated by
the coalescence of overlapping \FCs\ occupying the  same region of
volume $\xi^{d}$.

\medskip We would like to acknowledge useful conversations with 
F.\ J.\ Alexander and  H.\ Gould. The work of M.\ A.\ and W.\ K.\
was supported by DOE DE-FG02-95ER14498, and that of J.\ B.\ R.\ and
J.\ S.\ S.\ M.\ by  DOE DE-FG03-95ER14499. This work, LA-UR-00-0740,
 was also partially supported by the  Department of Energy under contract
W-7405.

\end{document}